\documentclass[12pt]{article}
\usepackage[dvips]{graphicx}

\def\ip{\lp \hangafter=1 \hangindent=20pt}      % hanging paragraph
\def\bib{\ip}                                   % hanging paragraph
\def\lp{\vskip 5pt \noindent} 
\def\la{\mathrel{\mathchoice {\vcenter{\offinterlineskip\halign{\hfil
$\displaystyle##$\hfil\cr<\cr\sim\cr}}}
{\vcenter{\offinterlineskip\halign{\hfil$\textstyle##$\hfil\cr
<\cr\sim\cr}}}
{\vcenter{\offinterlineskip\halign{\hfil$\scriptstyle##$\hfil\cr
<\cr\sim\cr}}}
{\vcenter{\offinterlineskip\halign{\hfil$\scriptscriptstyle##$\hfil\cr
<\cr\sim\cr}}}}}

\def\ga{\mathrel{\mathchoice {\vcenter{\offinterlineskip\halign{\hfil
$\displaystyle##$\hfil\cr>\cr\sim\cr}}}
{\vcenter{\offinterlineskip\halign{\hfil$\textstyle##$\hfil\cr
>\cr\sim\cr}}}
{\vcenter{\offinterlineskip\halign{\hfil$\scriptstyle##$\hfil\cr
>\cr\sim\cr}}}
{\vcenter{\offinterlineskip\halign{\hfil$\scriptscriptstyle##$\hfil\cr
>\cr\sim\cr}}}}}

\def\cmsq  {$\hbox{{\rm cm}}^{-2}$}    %cm-2
\def\percc {$\hbox{{\rm cm}}^{-3}$}    %cm-3
\def\MOLH {\hbox{${\rm H}_2$}}  %H2
\def\MOLN {\hbox{${\rm N}_2$}}  % N2
\def\AMM {\hbox{${\rm NH}_{3}$}} %NH3
\def\HCOP {\hbox{${\rm HCO}^+$}}      %HCO+
\def\HTHCOP {\hbox{${\rm H^{13}CO}^+$}}      %H13CO+
\def\HCEIOP {\hbox{${\rm HC^{18}O}^+$}}      %HC18O+
\def\HCSEOP {\hbox{${\rm HC^{17}O}^+$}}      %HC17O+
\def\DCOP {\hbox{${\rm DCO}^+$}}    %DCO+
\def\DTHCOP {\hbox{${\rm D^{13}CO}^+$}}    %D13CO+
\def\HTHP {\hbox{${\rm H}_{3}^{+}$}}   %H3+
\def\HTDP {\hbox{${\rm H}_{2}{\rm D}^{+}$}}   %H2D+
\def\HTHOP {\hbox{${\rm H}_{3}{\rm O}^{+}$}} % H3O+
\def\CEIO {\hbox{${\rm C}^{18}{\rm O}$}}   %C18O
\def\CSEO {\hbox{${\rm C}^{17}{\rm O}$}}   % C17O
\def\NTHP {\hbox{${\rm N}_2{\rm H}^+$}} % N2H+
\def\NTDP {\hbox{${\rm N}_2{\rm D}^+$}} % N2D+
\begin{document}

\begin{center}
\large

{\bf Deuterated molecules as a probe of ionization fraction in dense
interstellar clouds} \\

\vspace*{0.5cm}

\normalsize

P. Caselli \\
Osservatorio Astrofisico di Arcetri, Largo E. Fermi 5, I-50125 Firenze,
Italy; caselli@arcetri.astro.it 

\end{center}

\abstract{
The ionization degree $x(e)$ (= $n(e)/n({\rm H_2})$, with $n(e)$ and
$n({\rm H_2})$ the electron and H$_2$ number density, respectively)
plays a key role in the chemical and dynamical evolution of interstellar
clouds.  Gas phase ion--molecule reactions are major chemical routes to
the formation of interstellar molecules.  The time scale for ambipolar
diffusion of neutrals across field lines is proportional to the ionization
degree, which therefore is a crucial parameter in determining the initial
conditions which precede the collapse to form a star.  A direct measure
of $x(e)$ is hindered by the difficulty of observing H$_3^+$ and
H$_3$O$^+$, two of the
most abundant molecular ions, and atomic species with low ionization
potentials, such as atomic carbon and metals, which may be the main
repositories of positive charge.

Deuterium fractionation in molecular ions, in particular \HCOP ,
has been extensively used to estimate the degree of ionization in
molecular clouds. This paper reviews recent work on ionization 
degree in homogeneous clouds.  
We will show that the $N(\DCOP)/N(\HCOP )$ column 
density ratio furnishes a measurement of $x(e)$ only in regions 
where CO is not significantly depleted, thus in the outer skirts of 
dense cloud cores.  To probe $x(e)$ deep inside the clouds, one 
has to gauge deuterium enhancement in molecular ions with parent 
species not affected by depletion (e.g. \NTHP ), and
rely on chemical models which take into account the cloud density 
structure.  Unlike $N(\DCOP)/N(\HCOP )$, the $N(\NTDP )/N(\NTHP )$ column 
density ratio is predicted to considerably increase with core evolution
(and/or the amount of CO depletion),
reaching large values ($\ga$ 0.2) in cloud cores on the verge of forming a
star.}
  
\vspace*{0.5cm}


\section{Introduction}

The ionization fraction $x(e)$ 
of a dense interstellar cloud plays a fundamental 
role in chemical reaction schemes for molecular formation (e.g. Herbst 
\& Klemperer 1973; Oppenheimer \& Dalgarno 1974).
A knowledge of $x(e)$ also furnishes information on the ionization 
rate and thus on the sources which heat and ionize the gas. Moreover, the degree 
of ionization controls the coupling of the cloud to the magnetic field, and
thus regulates the rate of star formation.  This is due to the process of 
ambipolar diffusion (Mestel \& Spitzer 1956), whereby
the neutral particles contract relative to the ionized component which is 
tied to the field lines.  If the magnetic field threading the dense gas is 
sufficiently large to prevent immediate collapse, ambipolar diffusion of 
neutrals across field lines can lead to a situation where a dense core
of gas is gravitationally unstable. The time scale for ambipolar diffusion is 
proportional to the ionization degree (e.g. Spitzer 1978; Shu et al. 1987). 
Therefore our understanding of the physical processes and the dynamics 
of interstellar clouds requires a knowledge of electron and ion abundances.

In the theoretical determination of the degree of ionization we run into 
several sources of uncertainty, including the poorly known cosmic ray flux
and metal depletion within the cores, the penetration of UV radiation into
regions of high visual extinction due to cloud inhomogeneities, and the 
proximity to strong X--ray emitters (Caselli 2000; Caselli \& Walmsley 2000).
A direct measure of electron abundances is hindered by the difficulty of 
observing \HTHP \ (McCall et al. 1999) and \HTHOP \ (Phillips et al. 1992; 
Goicoechea \& Cernicharo 2001), two of the most abundant molecular ions in 
dense clouds, and atomic species with low ionization potential, such as 
atomic carbon and metals, which may be the main repositories of positive 
charge.  Rather, indirect determination of $x(e)$ through the estimate of 
molecular column densities sensitive to electron densities and the application
of chemical models are used.   

Observations of deuterated molecular ions 
and deuterium fractionation estimates represent a powerful tool to measure the 
ionization degree (Wootten et al. 1979; Gu\'elin et al. 1982; Dalgarno \& 
Lepp 1984; Caselli et al. 1998; Williams et al. 1998). 
In fact, using a simple steady state chemical model it is easy to show that
the [DCO$^+$]/[HCO$^+$] abundance ratio is inversely
proportional to $x(e)$ (Sect.~\ref{s-homo}).
Therefore it is tempting to use the observed $N({\rm DCO^+})/N({\rm HCO^+})$
column density ratio to infer  $x(e)$.  However, there are several caveats
to this simple picture, including a varying amount of depletion and molecular 
abundances along the line of sight (Sect.~\ref{s-homo}), and the 
existence of sharp density 
and temperature gradients (Sect.~\ref{s-grad}).  
We will discuss these caveats and show that in regions of high CO
depletion $N({\rm DCO^+})/N({\rm HCO^+})$ cannot be used to estimate
$x(e)$, because it is tracing conditions outside the depleted core.
Instead, other molecular ion column density ratios of the deuterated
to nondeuterated form (in particular $N({\rm N_2D^+})/N({\rm N_2H^+})$),
help in better constraining the ionization fraction in high density
regions, where depletion is more effective. 

\section{Electron fraction estimates in homogeneous clouds} 
\label{s-homo}

The formation of deuterated molecules starts with the isotope exchange 
reaction (Watson 1976):
\begin{eqnarray}
\HTHP + {\rm HD} & \rightleftharpoons & \HTDP + \MOLH , 
\label{ehd}
\end{eqnarray}
which is exothermic by an amount $\Delta E/k$ = 230 K (e.g. Millar et al. 1989).  
At the low temperatures typical of dense interstellar clouds,
the reverse of reaction (\ref{ehd}) is inhibited and the [\HTDP]/[\HTHP] abundance
ratio becomes much larger than the interstellar [D]/[H] ratio of 
1.5$\times$10$^{-5}$ (Linsky et al. 1995). The deuterium enhancement in \HTHP \
is limited by the rate at which electrons and neutrals destroy \HTDP .  
Any increase of the \HTDP /\HTHP \ abundance ratio due to reaction 
(\ref{ehd}) is reflected in other species formed through reactions with 
\HTHP , such as \HCOP . 
This can be easily shown by 
considering a simple chemical model with eight ingredients 
(H$_2$, CO, HD, H$_3^+$, \HTDP , \DCOP , \HCOP , and electrons).
  Besides reaction (\ref{ehd}), the following scheme is operative:
\begin{eqnarray}
%\MOLH + c.r. & \rightarrow & \MOLH^+ + e + c.r. \label{eh2} \\
%\MOLH^+ + \MOLH & \rightarrow & \HTHP + H \label{eh3p} \\
\HTHP + {\rm CO} & \rightarrow & \HCOP + \MOLH \label{ehcop} \\
\HTDP + {\rm CO} & \rightarrow & \DCOP + \MOLH  \label{edcop} \\
\HTHP + e & \rightarrow & products \label{eh3pe} \\
\HTDP + e & \rightarrow & products \label{eh2dpe} \\
\HCOP + e & \rightarrow & {\rm CO} + {\rm H} \label{ehcope} \\
\DCOP + e & \rightarrow & {\rm CO} + {\rm D} . \label{edcope} 
\end{eqnarray}
The rate equations governing the abundance of \HTDP ($n(\HTDP )$), 
\DCOP ($n(\DCOP)$) \ and \HCOP \ ($n(\HCOP )$) are: 
\begin{eqnarray}
\frac{dn(\HTDP )}{dt} & = & k_{\ref{ehd}} n(\HTHP ) n({\rm HD}) - k_{\ref{edcop}} 
n(\HTDP ) n({\rm CO}) - \beta_{\ref{eh2dpe}} n(\HTDP ) x(e) \\ 
\frac{dn(\DCOP)}{dt} & = & k_{\ref{edcop}} n(\HTDP) n({\rm CO}) - 
\beta_{\ref{edcope}} n(\DCOP) n(e) \label{edcopr} \\
\frac{dn(\HCOP)}{dt} & = & k_{\ref{ehcop}} n(\HTHP) n({\rm CO}) - 
\beta_{\ref{ehcope}} n(\HCOP) n(e) \label{ehcopr} 
\end{eqnarray}
At steady state, 
\begin{eqnarray}
n(\HTDP ) & = & \frac{k_{\ref{ehd}} n(\HTHP ) n({\rm HD})}{k_{\ref{edcop}}
n({\rm CO}) + \beta_{\ref{eh2dpe}} n(e)} \\
n(\DCOP ) & = & \frac{k_{\ref{edcop}} n(\HTDP) n({\rm CO})}{\beta_{\ref{edcope}} 
n(e)} \\
n(\HCOP ) & = & \frac{k_{\ref{ehcop}} n(\HTHP) n({\rm CO})}{\beta_{\ref{ehcope}} 
n(e)} \label{e-hcopsteady} ,
\end{eqnarray}
so that, assuming $\beta_{\ref{edcope}}$ = $\beta_{\ref{ehcope}}$ (Lee et al.
1996) and $k_{\ref{edcop}}$ = 1/3 $k_{\ref{ehcop}}$ (because \HTDP \ can also 
transfer one of the two protons to CO, thus producing \HCOP \ at a rate 
two times larger than that for \DCOP \ production),
\begin{eqnarray}
R_{\rm D} \equiv 
\frac{n(\DCOP )}{n(\HCOP )} & = & \frac{1}{3} \frac{n(\HTDP )}{n(\HTHP )} =
\frac{1}{3} \frac{k_{\ref{ehd}} n({\rm HD})}{k_{\ref{edcop}}
n({\rm CO}) + \beta_{\ref{eh2dpe}} n(e)} \label{erd} .
\end{eqnarray}
Substituting numerical factors in eqn.~\ref{erd} (see Caselli et al. 1998), 
assuming a kinetic temperature $T$ = 10 K and a CO fractional abundance
$x({\rm CO})$ = 9.5$\times$10$^{-5}$ (Frerking et al. 1982), 
we obtain a simple analytical expression which allow us to directly estimate 
$x(e)$ (= $n(e)/n(\MOLH )$) once $R_{\rm D}$ is known from observations:
\begin{eqnarray}
\frac{x(e)}{10^{-7}} & \simeq & \frac{0.27}{R_{\rm D}} - 1.9 \label{exe}.
\end{eqnarray}

Equation (\ref{exe}), obtained for illustrative purposes,  
is a very rough way to estimate the electron fraction in 
dense clouds, because several important ``complications'' have been neglected 
in deriving it.  First of all, eqn. (\ref{exe}) does not take into account 
the presence in 
the gas phase of other neutral species, besides CO, which contribute to the 
destruction of \HTDP .  In particular, atomic oxygen is predicted to be 
abundant in the gas phase (e.g. Lee et al. 1996; Caselli et al. 1998, 2001b) 
and recent ISO observations have confirmed this (e.g. Caux et al. 1999). This
has the effect of increasing the second term in the right hand side of 
eqn. (\ref{exe}) by a factor of a few.  Secondly, infrared and millimeter 
observations towards dense and cold clouds have established that a 
significant fraction of gaseous CO condenses onto dust grain surfaces
(Sect.~\ref{s-depl}), so that in eqn. (\ref{exe}) one should use 
$x({\rm CO})/f_{\rm D}$, where 1/$f_{\rm D}$ is the
fraction of CO in the gas phase.  The second term in eqn. (\ref{exe}) should 
thus explicitly include $f_{\rm D}$ and $x({\rm O})$, e.g.:
\begin{eqnarray}
\frac{x(e)}{10^{-7}} & \simeq & \frac{0.27}{R_{\rm D}} - 1.9 
\left[ \frac{1}{f_{\rm D}} + \frac{x({\rm O})/f_{\rm D}^{\prime}}{10^{-4}} 
\right] \label{exep}.
\end{eqnarray}
The depletion factor of atomic oxygen, $f_{\rm D}^{\prime}$, appears to be
significantly less than $f_{\rm D}$ (in the center of L1544, $f_{\rm D}^{\prime}$
$\sim$ 2--3, whereas $f_{\rm D}$ $\sim$ 1000; Caselli et al. 2001b), 
although direct observational determinations are not easy (e.g. Caux et al. 
1999).  Another problem with eqns. (\ref{exe}) and (\ref{exep}) is 
that they only furnish an average value of $x(e)$ along the line of sight,
with no clues on the variations of $x(e)$ inside the core which requires
a knowledge of the density and abundance gradients (Sect.~\ref{s-grad}). 
Finally, several simplifications have been made in the chemical scheme
(\ref{ehd}) - (\ref{edcope}), by neglecting important ingredients in the 
electron balance of dense clouds, including recombination of atomic and 
molecular ions on negatively charged dust grains (Draine \& Sutin 1987), 
refractory metals (although significantly depleted; see Sect.~\ref{s-l1544}),
and the presence of 
atomic deuterium (Dalgarno \& Lepp 1984) which somewhat
increases the deuterium fractionation (Anderson et al. 1999).

\subsection{The effects of depletion on deuterium fractionation and $x(e)$}
\label{s-depl}

 For a normal dust to gas ratio and assuming typical dust 
grains with radius of 0.1 $\mu$m, the time scale for a gaseous species 
to be deposited onto a grain is $t_{\rm D}$ $\sim$ 10$^9 
\sqrt{A_{\rm X}}/[S n(\MOLH )]$ yr, 
where $A_{\rm X}$ is the molecular weight, $S$ is the sticking coefficient and 
$n(\MOLH )$ is the number density of molecular hydrogen (e.g. Burke \& 
Hollenbach 1983; van Dishoeck et al. 1993).  For unity sticking coefficient
(Jones \& Williams 1985; Tielens \& Allamandola 1987) and a typical 
\MOLH \ number density of 10$^4$--10$^5$ \percc ,  $t_{\rm D}$ $\sim$ 
10$^5$ -- 10$^4$ yr, comparable to the dynamical time scale of dense cores.
Thus, gas--phase depletion is expected to occur in molecular clouds.

Infrared and millimeter observations have indeed revealed large amounts of 
CO depletion.  In quiescent regions of the Taurus molecular cloud, up to 
40\% of the total (gas+ice) CO is found in solid form (Chiar et al. 1995).
In the densest parts of the IC 5146 molecular cloud, Kramer et al. (1999) 
found a systematic fall off in the ratio between the \CEIO \ column density 
and the visual extinction, consistent with CO depletion.  Large depletion 
factors ($f_{\rm D}$ $\ga$ 10) have been measured towards well known 
pre--stellar cores: L1544 (Caselli et al. 1999; Sect.~\ref{s-l1544}), 
L1498 (Willacy et al. 1998; Tafalla et al. 2001),
L1698B (Jessop \& Ward--Thompson 2001), L1495, L1400K, and L1517B (Tafalla et 
al. 2001).

Following Dalgarno \& Lepp (1984), Caselli et al. (1998) have shown that 
uncertainties on the amount of neutral depletion implies an
ambiguous determination of the electron fraction (see their eqn. (3) and 
Tab.~7).  Thus, it is extremely important to estimate $f_{\rm D}$ from
observations before attempting any $x(e)$ measurement.  In fact,
decreased fractional
abundances of CO and other neutral species causes a drop in
the \HTDP \ and \HTHP \ destruction rates, and a rise in the \HTDP \ abundance
(via reaction ~\ref{ehd}) due to the larger \HTHP \ abundance.  The net result
is more efficient deuterium fractionation (see eq.~\ref{erd}).  
Large values in the observed 
$R_{\rm D}$ and the use of eqn. (\ref{exe}),
which does not account for CO depletion, will give erroneously low $x(e)$ 
values (note that eqn. (\ref{exe}) cannot account for $R_{\rm D}$ $\geq$ 0.07).

To see how the electron fraction is affected by the depletion of 
neutrals, we consider a ``zero--order'' analysis, using 
eqn. (\ref{e-hcopsteady}), generalized to a generic neutral species 
$m$ ($x(m)$ $\simeq$ $x({\rm CO})$ + $x({\rm O})$):
\begin{eqnarray}
x(m{\rm H}^+ ) & = & \frac{k_m x(\HTHP) x(m)}{\beta_{\ref{ehcope}} 
x(e)} , \label{exm}
\end{eqnarray}
and the steady state equation for $x(\HTHP )$:
\begin{eqnarray}
x(\HTHP ) & = & \frac{\zeta / n(\MOLH )}{k_m x(m) + 
\beta_{\ref{eh3pe}} x(e)} \label{eh3p},
\end{eqnarray}
where $\zeta$ is the cosmic ray ionization rate, $n(\MOLH )$ is the 
\MOLH \ number density, and $k_m$ is the generic rate coefficient for \HTHP \
destruction reactions with $m$ ($k_m$ $\sim$ 10$^{-9}$ cm$^{3}$ s$^{-1}$). 
Substituting eqn. (\ref{eh3p}) into 
eqn. (\ref{exm}) and solving for $x(e)$, we arrive at a 
quadratic equation with solution:
\begin{eqnarray}
x(e) & = & -q \, x({m}) [1 - \sqrt(1+z/x(m))] \label{esol} \\
q & = & \frac{k_m}{2 \beta_{\ref{eh3pe}}} \\
z & = & \frac{4 \beta_{\ref{eh3pe}} \zeta /n(\MOLH )}{\beta_{\ref{ehcope}} 
k_m x(m{\rm H}^+)} \label{ezeta}. 
\end{eqnarray}
The square root term in the right hand side of eqn. (\ref{esol})
can be expanded in a binomial series, which, truncated at the first order term,
gives:
\begin{eqnarray}
x(e) & = & q \, z /2 
\end{eqnarray}
independent of $x(m)$, but inversely dependent on $x(m{\rm H^+})$ (see 
eqn.~\ref{ezeta}).  This means that the electron fraction tends to 
increase with freezing out of neutrals, if the amount of $m$ depletion 
becomes large enough to reduce the $m$H$^+$ formation rate, despite of the 
increased \HTHP \ abundance.  However, even if CO is completely depleted
(see Sect.~\ref{s-homo}), atomic oxygen may maintain a considerable 
gas phase abundance even at large densities ($n(\MOLH )$ $\ga$ 10$^6$ \percc ;
see Sect.~\ref{s-l1544}).  Thus, in zero--order, 
CO depletion {\it alone} will not significantly affect $x(e)$.  As we will
see in Sect.~\ref{s-grad}, estimates of $f_{\rm D}$ are needed anyway to 
fine--tune chemical models.

\subsection{The use of detailed chemical models}
\label{s-detailed}

There have been several attempts in the past to use either the observed 
degree of deuterium fractionation (e.g. Gu\'elin et al. 1982; Dalgarno \& 
Lepp 1984) 
or the abundance of selected molecular ions (de Boisanger et al. 1996),
together with chemical models, 
to determine the electron fraction.
Recent improvements in the laboratory measurements of the \HTHP \ and 
\HTDP \ dissociative recombination rates (Larsson et al. 1996; Kokoouline
et al. 2001),
crucial in chemical models, as well as the increased sensitivity, spectral 
and angular resolution of millimeter telescopes, have furnished new impetus
to refine previous determinations of $x(e)$.  
 
Caselli et al. (1998) used the ``New Standard Model'' of Lee et al. (1996),
modified in order to include simple deuterated species, to reproduce the 
observed $x(\DCOP )/x(\HCOP )$ ($\equiv$ $R_{\rm D}$) and 
$x(\HCOP )/x({\rm CO})$ ($\equiv$ $R_{\rm H}$) abundance ratios 
and thus to put constraints on $x(e)$ in a sample of low mass cloud cores 
studied by Butner et al. (1995) with the MWO 4.9m telescope (Texas) and the 
NRAO 12m antenna.  Free parameters in these models are: (i) 
the depletion factor, $f_{\rm D}$; (ii) the abundance of refractory metals,
such as Fe, Mg, and Na, initially present in singly charged atomic form, 
$x(M^+)$; (iii) the cosmic ray ionization rate, $\zeta$; and (iv) $n(\MOLH )$.
Whereas $f_{\rm D}$ and $n(\MOLH )$ can be estimated from observations, 
the other two parameters ($x(M^+)$ and $\zeta$) 
have to be determined through the comparison
between observed quantities and model results, in the same manner as
$x(e)$.  To better constrain these quantities, two observables ($R_{\rm D}$ 
and $R_{\rm H}$) were
 included in the analysis.  $x(M^+)$ and $\zeta$ were varied to cover
the range of observed abundance ratios.  Using a least--square method to 
model--fit the
 calculated data points, Caselli et al. (1998) found that typical 
parameters in low mass cloud cores are: 
\begin{eqnarray*}
10^{-8} & \la x(e) \la & 10^{-6} \\
10^{-18} & \la \zeta ({\rm s}^{-1}) \la & 10^{-16} \\
10^{-10} & \la x(M^+) \la & 10^{-7},  
\end{eqnarray*}
with $f_{\rm D}$ between 2 and 5.  We note that, although $x(M^+)$ is  
poorly constrained, the values obtained by Caselli et al. (1998) 
corroborate the view of Graedel et al. (1982) that metals are highly depleted 
in dense cloud cores. This is indirectly confirmed by the fact that molecules
containing Fe, Mg, and Na have not been discovered in molecular clouds.
 Given the assumed homogeneity of the dark cloud
cores, the deduced ranges in $x(e)$ and $\zeta$ may appear too large, 
unless ionization losses are significant through modest column densities of 
material.  Indeed, theoretical (e.g. Hartquist \& Morfill 1983, 1994) and 
experimental (Fukui \& Hayakawa 1981) work have shown that this is possible
(see also Caselli 2000).  

Similar $x(e)$ values ($10^{-7.5} \la x(e) \la 10^{-6.5}$) have been 
deduced in another sample of 
low mass cores by Williams et al. (1998), using the  Bergin 
\& Langer (1997) chemical model
and new observations of CO, \HTHCOP , and \DCOP \ at the 
NRAO 12m antenna (in this case, $\zeta$ was fixed at 5$\times$10$^{-17}$ 
s$^{-1}$ and the depletion factor was varied between 2 and
0.8). Applying the same method to more massive cores, Bergin et al. (1999) 
found electron abundances within a smaller range,  $10^{-7.3}  \la x(e) \la 
 10^{-6.9}$.  Only 
the most massive sources stand out as having the lowest electron 
abundances, $x(e)$ $<$ 10$^{-8}$, in agreement with the findings of 
de Boisanger et al. (1996).

\section{$x(e)$ in heterogeneous clouds }
\label{s-grad}

Dense molecular clouds are heterogeneous and highly clumped. A 
knowledge of the cloud density
structure is necessary to accurately estimate the electron fraction across 
the core.  From submillimeter continuum dust emission, Ward--Thompson et al.
(1994) found that the radial density profiles of starless cores, assuming
a constant dust temperature, is $n(r)$ $\propto$ $r^{-\alpha}$, with 
$\alpha$ $\sim$ 0.4--1.2 at radii less than
$\sim$ 4000 AU (the so--called ``flattened region''), and approaches 
$n(r)$ $\sim$ $r^{-2}$ between $\sim$
4000 AU and $\sim$ 15000 AU (Andr\'e et al. 1996, Ward--Thompson et al. 1999).
This result has been recently questioned based on 
model calculations of the dust
temperature in pre--stellar cores (Zucconi et al. 2001; Evans et al. 2001),
which predict temperature drops from  about 14 K at the outer core edges  
to $\sim$ 7 K at the center.  Temperature gradients imply more peaked
density profiles, with smaller ``flattened'' zones and larger central 
densities, similar to the density structure of 
cores with stars,  typically consistent 
 with power--laws of the form $n(r)$ $\propto$ $r^{-2}$
or $r^{-1.5}$ (Motte \& Andr\'e 2001).     

The electron fraction estimates listed in Sect.~\ref{s-homo} are based on
data with angular resolution of the order of an arc minute, which corresponds
to about 0.04 pc at the distance of Taurus and Ophiuchus, where most of the 
studied cores are located.  As a consequence, the density profile cannot be 
resolved, at least in the central part of the core.  To study spatial 
variations in the ionization degree of a dense cloud core one needs to map 
(i) the dust continuum emission, to derive the density profile, and (ii) 
several species (at least CO, \HCOP \ and \DCOP , but, as shown in 
Sect.~\ref{s-l1544}, this is not enough) using larger telescopes.  
So far, a single detailed 
study of the electron fraction variation in dense cores has been performed 
(in L1544; Caselli et al. 2001b) and in the next section we will summarize the 
observations and the data analysis which have been used to estimate 
$x(e)$ as a function of cloud radius.

\subsection{The starless core L1544}
\label{s-l1544}

\subsubsection{Observational results}

L1544, a starless core in the Taurus Molecular Cloud, is thought to be 
in an early stage of the star formation process, with evidences for
gravitational collapse and no signs of infrared point sources, indicative
of the presence of young stellar objects.  Thus, it is a 
good target to investigate the initial conditions for stellar 
birth.  L1544 has been extensively 
studied in several high density tracers such as CS, CCS, \NTHP  (Tafalla 
et al. 1998; Williams et al. 1999; Ohashi et al. 1999) and \CSEO \ (Caselli et 
al. 1999).  Important results from these studies indicate that (a) L1544 
is surrounded by a low density envelope which is undergoing extended infall,
and (b) CO is highly depleted at densities above $\sim$ 10$^5$ \percc , or 
inside radii of $\sim$ 6000 AU.  The high density core nucleus was however 
not studied in detail because the observed lines avoided the central region
due to freeze out of the selected molecules onto dust grains (as in the 
case of CO, CS, and CCS), or because the low J transitions 
(e.g. J = 1--0 for \NTHP ) observed are affected by absorption in 
the low density foreground material.

\begin{figure}{}
%\vspace*{-4cm}
%\resizebox{10cm}{!}{\includegraphics[angle=-90,width=20.0cm]{casellip1.ps}}
\centerline{\includegraphics[angle=-90,width=15.0cm]{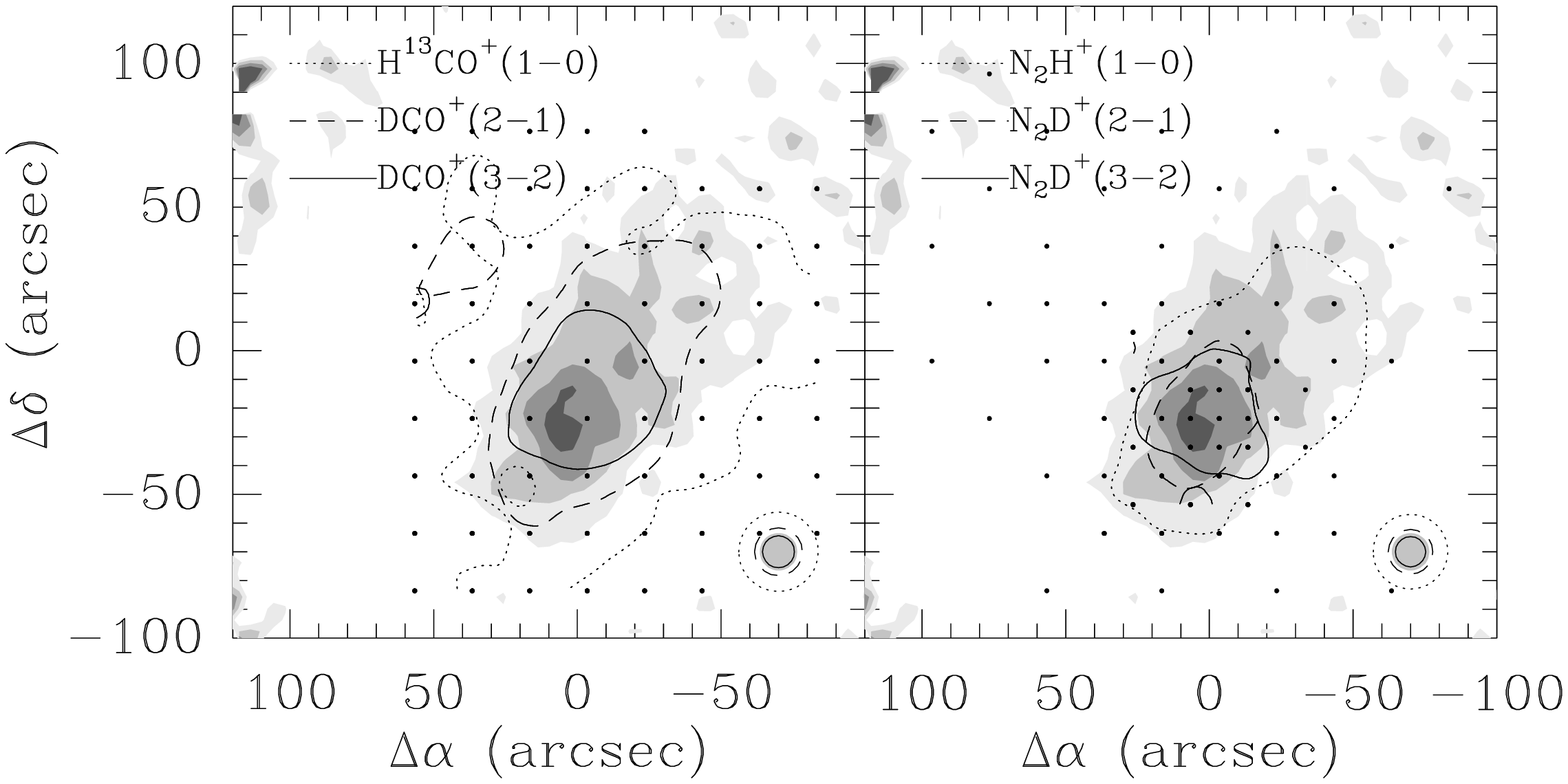}}
%\resizebox{10cm}{!}{\includegraphics{integrated_sminiato.ps}}
\vspace*{2cm}
\caption{
\footnotesize
Half maximum contours of integrated intensity maps of ({\it left}) 
\HTHCOP (1--0) (dotted contour), \DCOP (2--1) (dashed), \DCOP (3--2) (thin), 
and ({\it right}) \NTHP (1--0) (dotted),
\NTDP (2--1) (dashed), and \NTDP (3--2) (thin), overlapped with the 1.3mm 
continuum dust emission from Ward--Thompson et al. (1999) (gray scale; 
see Caselli et al. 2001a for details on the
observations).  Beam sizes at the corresponding frequencies are reported in the
bottom right corner.  \HCOP \ and \DCOP \ are more extended than \NTHP \ and 
\NTDP , due to the differential depletion of CO and \MOLN , which 
causes strong abundance gradients in the core (see text).}
\label{fig1}
\end{figure}

We have recently undertaken an extensive survey of molecular ions which 
allowed us to make detailed maps of 
the L1544 nucleus and investigate gas kinematics 
as well as determining the electron fraction across the core (Caselli et al.
2001a, 2001b).  We mapped \HTHCOP (1--0), \DCOP (2--1), \DCOP (3--2), 
\NTHP (1--0), \NTDP (2--1), and \NTDP (3--2) using the IRAM 30m antenna at 
Pico Veleta.  Unlike the CO 
emission,  all the molecular ions trace the dust distribution approximately, 
indicating that depletion is not so severe as in the case of carbon monoxide 
(Caselli et al. 1999).  There are however differences between different 
tracers and 
this is evident in Fig.\ref{fig1},  where the 
half maximum contours of \HCOP \ and \NTHP \ isotopomers are compared.  
\HTHCOP \ and \DCOP \ are more extended than \NTHP \ and \NTDP.  Moreover, 
\HCOP \ and \DCOP \ peak slightly but significantly off the dust peak, whereas 
\NTHP \ follows the dust continuum emission and \NTDP \  
traces the {\it core nucleus}, the future stellar cradle.

The formyl ion,  formed through the reaction between CO and \HTHP ,  
can be considered a good CO tracer in high density gas because it has a higher 
dipole moment than the parent species.  Analogously, \NTHP \ can be considered 
a good tracer of the unobservable \MOLN \ .  From the 
morphological differences between the various molecular ions shown in 
Fig.~\ref{fig1},  we deduce that CO is less volatile than \MOLN, 
and, consequently, \HCOP \ and \DCOP \ are less abundant in the core nucleus 
than \NTHP \ and \NTDP . This confirms the conclusions of Caselli et al. 1999
 who found large degrees of CO depletion towards the L1544 dust peak position
and suggests the presence of strong abundance gradients in the core.  In 
particular, toward the dust emission peak, we found different degrees of 
deuterium fractionation in 
\HCOP \ and \NTHP \ ($N(\DCOP )/N(\HCOP )$ = 0.04$\pm$0.02;  
$N(\NTDP )/N(\NTHP )$ = 0.24$\pm$0.02), which can be reproduced only if 
differential
depletion of molecular species onto dust grains, and density gradients are 
taken into account in chemical models (models without depletion gradients 
predict similar degrees of deuterium fractionation in \HCOP \ and \NTHP;  e.g. 
Rodgers \& Charnley 2001).  As discussed in Caselli et al. (2001b), the 
greater volatility of
 \MOLN \ allows \NTHP \ to be more abundant than \HCOP \ in the
dense and highly CO--depleted central regions of the core.  Given that 
depletion favors deuterium fractionation (see Sect.~\ref{s-depl}), the higher
$N(\NTDP )/N(\NTHP )$ column density ratio is thus naturally explained as 
arising from an inner and denser region than that probed by 
$N(\DCOP )/N(\HCOP )$.  From this, one can conclude that $x(e)$ measurements 
in the center of dense cloud cores, where CO is highly depleted,  cannot be 
based on \HCOP \ and \DCOP \ observations alone.

\subsubsection{Chemical model results}
\label{s-model}

We approximate L1544 by a spherically symmetric cloud with a density profile
deduced from 1.3mm continuum dust emission observations (Ward--Thompson et al.
1999): a central density of about 10$^6$ \percc \ inside a radius of 2500 AU, 
followed by a $r^{-2}$ density fall off until a radius of 10000 AU.  We 
also assumed isothermal conditions with $T$ = 10 K, based on recent 
observations of \AMM (1,1) and (2,2) with the Effelsberg telescope (Tafalla
et al. 2001). The temperature drop to 7 K in the central 0.01 pc, 
predicted by Evans et al. (2001) and Zucconi et al. (2001), 
which is probably not 
resolved by the above mentioned \AMM \ observations, does not significantly 
affect our chemical results.   Neutral
species in the model are \MOLH , CO, \MOLN , and atomic oxygen. A fundamental
assumption here is that all gas phase N is in molecular form, and C is in CO.
This assumption is validated by the general agreement with the ``late time''
results of the detailed chemical--dynamical model by Aikawa et al. (2001) 
applied to L1544.  We follow 
depletion of these species onto dust grains and their desorption due to 
cosmic ray impulsive heating of the dust, following the procedure of 
Hasegawa \& Herbst (1993).  The abundance of  \HCOP and \NTHP \  are given 
by the steady state chemical equations using the instantaneous abundances of 
neutral species.  Analogously, the electron fraction $x(e)$ is
computed in terms of global estimates for the molecular and metallic ions 
and using instantaneous abundances of CO, \MOLN , and O in the gas phase, 
using a simplified version of the reaction scheme of Umebayashi \& Nakano 
(1990).  

The $x(\DCOP )/x(\HCOP )$ and $x(\NTDP )/x(\NTHP )$ abundance
ratios, $R_{\rm D}$,  are computed assuming that the deuterated species 
forms by proton transfer from \HTDP :
\begin{eqnarray}
R_{\rm D} & = & \frac{k_{\ref{ehd}} x({\rm HD})}{3 [\beta_{\ref{eh2dpe}}x(e) + 
k_m x(m) + k_g x(g)]} . \label{erdl1544} 
\end{eqnarray}
In the above equation, $k_{\ref{ehd}}$ is the rate coefficient of 
reaction \ref{ehd}, $\beta_{\ref{eh2dpe}}$ is the rate for dissociative 
recombination of \HTDP \ (reaction \ref{eh2dpe}), 
$k_m$ is the rate coefficient for \HTHP \ (and 
\HTDP ) destruction with neutral species $m$, and $k_g$ is the rate for 
recombination of \HTDP \ on grain surfaces (Draine \& Sutin 1987).  
$x(m)$ is an average over CO, \MOLN, and O abundances, weighted by the 
respective depletion factors, and $x(g)$ is the grain abundance by number
(computed, together with $k_g$, using the MNR (Mathis et al. 1977) grain size 
distribution).  

We run several models changing parameters such as the cosmic ray ionization 
rate ($\zeta$), the CO, \MOLN , and O binding energies onto grain surfaces 
($E_{\rm D}$), the lower cut--off radius ($a_{\rm min}$) in the MRN grain size
distribution, and the sticking coefficient ($S$) which 
gives the probability for a species colliding with a grain to get adsorbed 
onto its surface.  The ``best fit'' model, which reproduces the molecular
column densities observed towards the L1544 dust peak and best follows the 
observed column density profiles (see Model 3 in Caselli et al. 2001b), has 
the following parameters:
$\zeta$ = 6$\times$10$^{-18}$ s$^{-1}$ (a factor of 2 smaller than the 
standard 
value, usually adopted in chemical models);  $S$ = 1 (see e.g. Jones \& 
Williams 1985);  $E_{\rm D}$(CO) = 1210 K, 
appropriate for polar mantles (Tielens \& Allamandola 1987);
$E_{\rm D}$(\MOLN ) = 787 K, 1.5 times smaller than the CO binding energy, 
as found in calculations by Sadlej et al. 1995;
$E_{\rm D}$(O) = 600 K, a best fit value (1.3 times smaller than that estimated
for polar mantles); $a_{\rm min}$ = 5$\times$10$^{-6}$ cm (5 times larger 
than the MRN value).  Whereas $E_{\rm D}$(CO), $E_{\rm D}$(\MOLN ), and 
$S$ have been fixed to values found in the literature, 
$\zeta$ and $E_{\rm D}$(O) were allowed to change and in fact they 
are the two crucial ``free'' parameters in the model.  
Assuming that the cosmic ray ionization rate does not change across the
cloud, a combination of moderately low $\zeta$ and O binding energy
is necessary to (i) match the observed \HCOP \ and \NTHP \ column 
density profiles, and (ii) keep the deuterium fractionation relatively low 
in the cloud center and reproduce the observed $N(\NTDP )/N(\NTHP )$ 
and $N(\DCOP )/N(\HCOP )$ column density ratios.  Good agreement with 
observations can also be reached if $\zeta$ is decreasing toward the core
center, and we are currently exploring this possibility.

\begin{figure}{}
\vspace*{-2cm}
\centerline{\includegraphics[angle=-90,width=15.0cm]{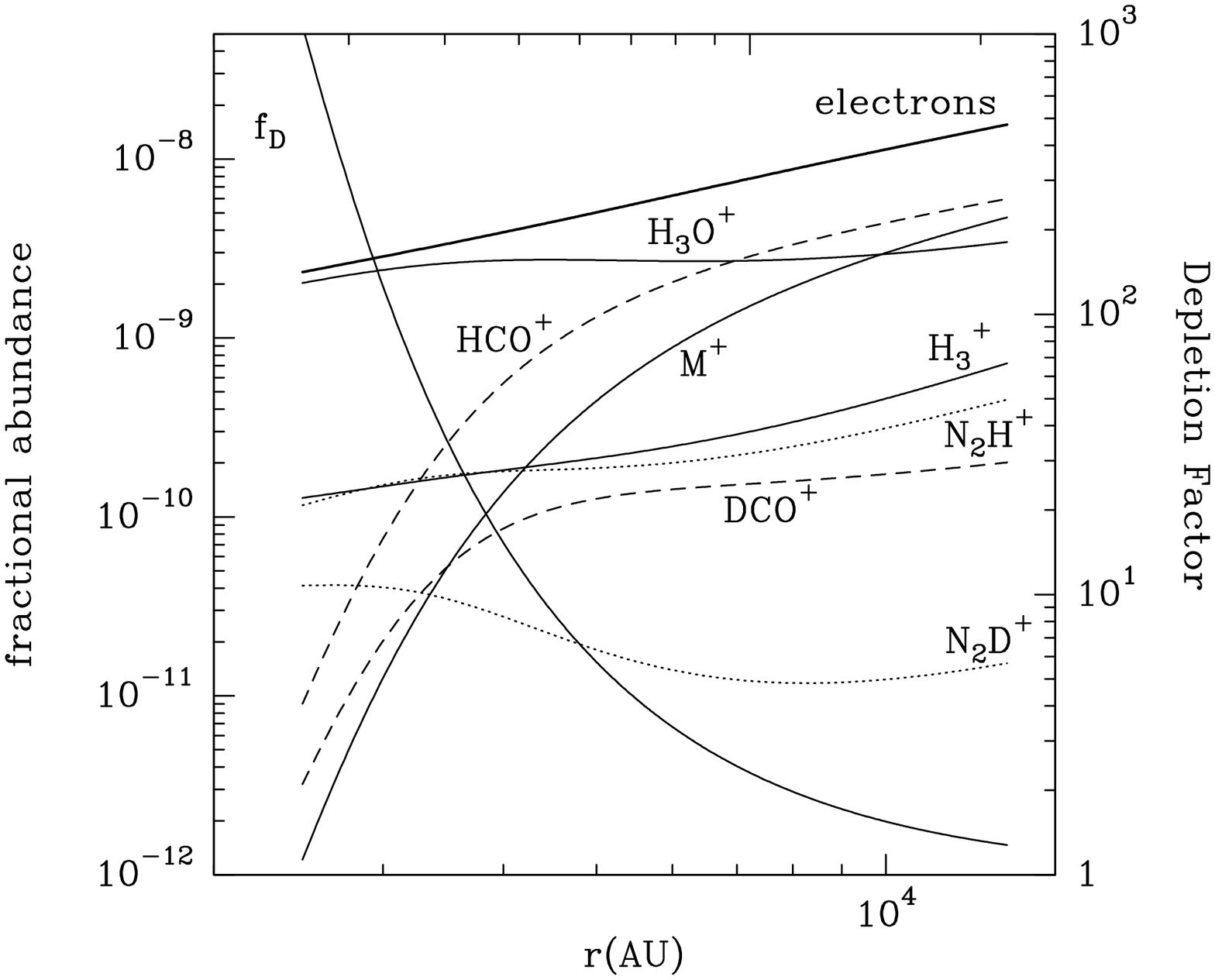}}
%\resizebox{10cm}{!}{\includegraphics{integrated_sminiato.ps}}
\vspace*{-1.5cm}
\caption{
\footnotesize
Radial profiles of fractional abundances
of electrons, \HCOP , \DCOP , \NTHP , \NTDP , \HTHOP , \HTHP \ and metal
ions (M$^+$)
predicted by the best fit model of Caselli et al. (2001b). $f_{\rm D}$ is  
the CO depletion factor.  \HCOP \ and \DCOP \ abundances drastically 
drop below $r$ $\sim$ 4000 AU, whereas no ``hole'' is present in the \NTDP \
profile.  Inside the flattened region,  the main molecular ion is \HTHOP .
The assumed density profile is based on 1.3mm continuum dust emission 
observations (Ward-Thompson et al. 1999).}
\label{fig2}
\end{figure}

Radial profiles of molecular and electron fractional abundances from 
$r$ $\sim$ 2500 AU to $r$ $\sim$ 10000 AU are shown in Fig.~\ref{fig2}.
Here, all the species contributing to the electron fraction are displayed,
unlike Fig.~8 of Caselli et al. (2001b), where only observed molecular
ions are shown.  Several interesting predictions can be based on  
these profiles: 
(i) \HCOP \ and \DCOP \ abundances rapidly
drop below 4000 AU; (ii) \HTHOP \ becomes the major molecular ion once
\HCOP \ starts to drop (in agreement with Aikawa et al. 2001); 
(iii) CO starts to be highly depleted inside 5000 AU 
(where the fraction of CO in the solid phase is about 1\% of the total 
``canonical'' CO abundance of 9.5$\times$10$^{-5}$ (Frerking et al. 1982)) 
and it is almost totally 
depleted inside 3000 AU (where 99\% of the available CO is frozen onto dust 
grains); (iv) \MOLN , being more volatile than CO, maintains large fractions
of \NTHP \ and \NTDP \ in the core center, although the \NTHP \ abundance curve
presents a slight fall off at $r$ $\la$ 3000 AU;  (v) large degrees of 
deuterium fractionation are present at the core center, where 
$x(\NTDP )/x(\NTHP )$ $\sim$ 0.4, due to the large degree of CO depletion; 
(vi) {\it electron fraction} values are a few
times 10$^{-9}$ toward the core center and increases to a few times 10$^{-8}$ 
at the outer edge of the core.  The 
central value of $x(e)$ is a factor of about 6 lower than that deduced 
from the standard relation between $x(e)$ and \MOLH \ number density
 ($x(e)$ $\sim$ 
1.3$\times$10$^{-5}$ $n({\rm H_2})^{-1/2}$; McKee 1989). The relation we find 
from the model is:
\begin{eqnarray}
 x(e)  =  5.2 \times 10^{-6} \times n(\rm H_2)^{-0.56} \label{exen}.  
\end{eqnarray}
Once the electron fraction is known and assuming a balance 
between the magnetic field force -- transmitted to neutrals by ion--neutral 
collisions -- and gravity, the 
ambipolar diffusion time scale is immediately deduced: $t_{\rm AD}$ = 
2.5$\times$10$^{13}$ $x(e)$ yr in the case of a cylinder (Spitzer 1978)
and $t_{\rm AD}$ = 3.7$\times$10$^{13}$ $x(e)$ yr for an oblate 
spheroid (McKee et al. 1993).  The interesting thing is that $t_{\rm AD}$ 
toward the core center is comparable (within a factor of 2--3) 
to the free--fall time scale, $t_{\rm ff}$, at densities of about 10$^6$ 
\percc , consistent with the idea that the nucleus of the L1544 core is 
close to dynamical collapse.  Similarly, Umebayashi \& Nakano (1990) found 
$t_{\rm AD}$/$t_{\rm ff}$ $\sim$ 4 in their case 2 (the closest to our model)
at  $n(\MOLH )$ = 10$^6$ \percc . 
Note that \HTHP \ closely follows \NTHP , whereas 
metal abundances, which are never primary repositories of positive charges 
in L1544, are negligible at the core center. The undepleted abundances 
of metals ($x_i(M)$ = 10$^{-7}$) are about one order of magnitude lower 
than those  measured in 
diffuse clouds (Morton 1974), as is usual in chemical models of dense clouds 
(e.g. Herbst \& Leung 1989; Graedel et al. 1982; Lee et al. 1986; see also 
Sect.~\ref{s-detailed}).  On the 
other hand, McKee (1989) adopts $x_i(M)$ = 10$^{-6}$ in modeling the 
ionization in (undepleted) molecular clouds.  
This leads to electron fractions about 
an order of magnitude larger than predicted by our model.

\subsection{Predictions for dense clouds}
\label{s-pred}

The chemical model described in Sect.~\ref{s-model} can be used to investigate
the distribution of electron fraction and molecular ions, including the 
unobservable \HTHP \ and \HTHOP , in cores with different masses, density 
profiles, and different amounts of CO depletion compared to L1544.  We will  
describe (i) the case of a ``pivotal core'',  more centrally concentrated and 
with larger central densities than L1544, which 
represents a more massive object ($M$ = 12 M$_{\odot}$, twice the L1544 mass 
deduced from the dust emission continuum flux), and perhaps a more evolved 
stage in the process of star formation; 
(ii) a ``flattened core'', with a total mass of $\sim$ 1 M$_{\odot}$, 
shallower density profile and smaller 
amount of CO depletion than L1544, representing a starless low mass core 
in an earlier phase of evolution.

\subsubsection{The ``pivotal core''}
\label{s-spiky}

 In the widely used ``standard'' model of isolated star formation (e.g. Shu 
et al. 1987), a singular isothermal sphere (SIS; Shu 1977) 
is considered to be the pivotal 
state (or starting point) for dynamical cloud collapse.
We expect a starless core such as L1544 to evolve toward 
an even more centrally concentrated density distribution.
In this section 
we analyse the chemical composition and the electron fraction of such 
a cloud, assuming a sphere with constant density of 10$^7$ \percc \ inside
$r_f$ = 1000 AU (instead of 2500 AU, as in L1544) and external radius at 
15000 AU (as in 
L1544).  The total mass of such an object is 12 M$_{\odot}$. 
The chemical parameters are the same as in the ``best 
fit'' model described in Sect.~\ref{s-model}. 

\begin{figure}{}
%\vspace*{-2cm}
\centerline{\includegraphics[angle=-90,width=15.0cm]{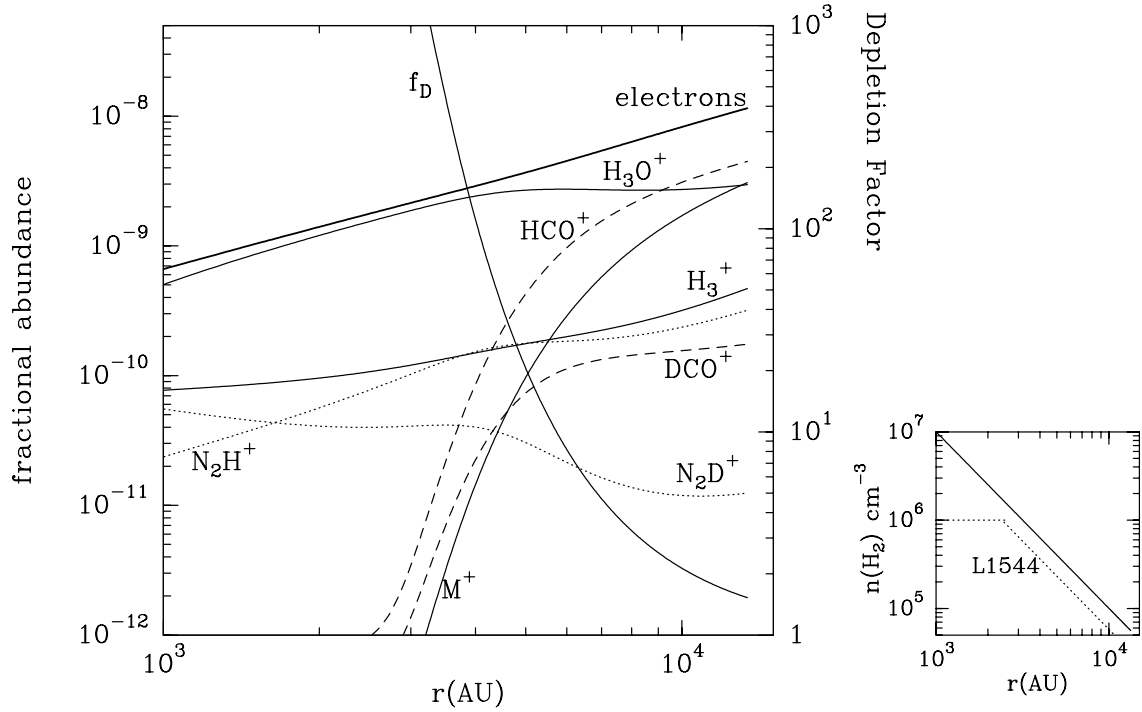}}
%\resizebox{10cm}{!}{\includegraphics{integrated_sminiato.ps}}
\vspace*{-1.5cm}
\caption{
\footnotesize
Same as Fig.~\ref{fig2} 
for a core more centrally concentrated (and eventually more evolved) than 
L1544.  The ``flattened'' region is 1000 AU in size and has a density of
10$^7$ \percc .  $f_{\rm D}$ is the CO depletion factor.  As in L1544 
(see Fig.~\ref{fig2}; note the different abscissa scale), \HCOP \ and \DCOP \ 
abundances drastically drop below $r$ $\sim$ 4000 AU.  \NTHP \ gradually 
declines toward the center, and becomes less abundant than \NTDP below 2000 
AU.  The main molecular ion is \HTHOP .  The figure on the right shows the 
density profile of the pivotal core (continuous line), compared to the L1544 
profile (dotted curve).}
\label{fig3}
\end{figure}

Results of this more centrally concentrated model are displayed 
in Fig.~\ref{fig3}.  The large amount of CO depletion inside 4000 AU produces
a sharp drop (or a ``hole'') in the  \HCOP \ and \DCOP \ abundance profiles. 
No ``hole'' is present in \NTDP , whereas the 
\NTHP \ abundance profile presents a central ``valley'' of radius $r$ $\sim$ 
3000 AU, 
with a much shallower decline than that of \HCOP \ and \DCOP . 
Inside the \NTHP \ valley, deuterium
fractionation becomes so important that [\NTDP ]/[\NTHP ] $\ga$ 1 
at 1000 AU.   \HTHP \ more or less follows the \NTHP \ abundance curve, within 
a factor of 2. Another thing to note is that \HTHOP \ is the main 
ion all across the core\footnote{If a larger binding enery for atomic 
oxygen is 
used ({\it e.g.} the typical 800 K, instead of 600 K adopted in our best fit 
model), \HCOP \ is the main ion until a radius 
of about 4000 AU.  At smaller radii,  \HTHOP , \HTHP , \NTHP \ and \NTDP \ 
almost equally contribute to the ionization degree.}. \HCOP \ becomes 
predominant only 
at the outer edges, where CO depletion is negligible, and where S$^+$,
C$^+$ and metal ions are also important repositories of positive charge.    
Therefore, \HCOP \ column densities and derived abundances can be used as
rough estimates of the 
electron fraction in the core envelope.  At smaller radii, 
one has to rely on \NTHP \ and \NTDP \ observations.
As in the case of L1544, the predicted central $x(e)$ value ($\sim$ 
7$\times$10$^{-10}$) implies an ambipolar diffusion time scale locally 
similar to the free--fall time scale (see Sect.~\ref{s-model}).

\subsubsection{Flattened cores}
\label{s-flat}

\begin{figure}{}
%\vspace*{-2cm}
\centerline{\includegraphics[angle=-90,width=15.0cm]{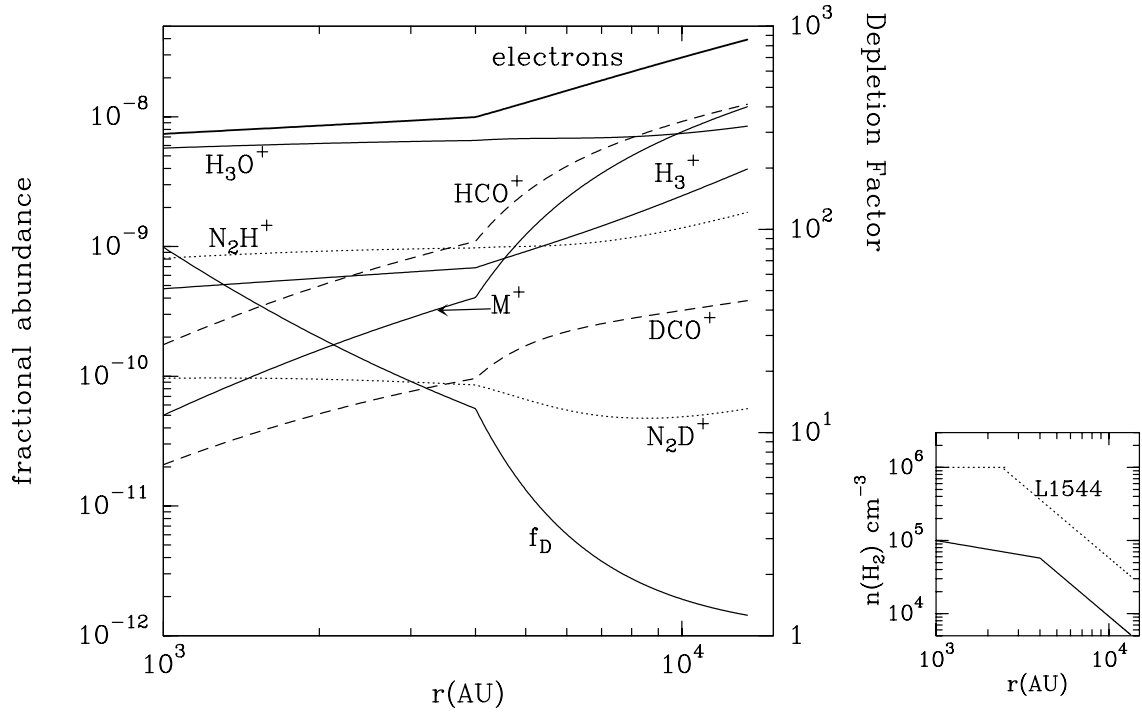}}
%\resizebox{10cm}{!}{\includegraphics{integrated_sminiato.ps}}
\vspace*{-1.5cm}
\caption{
\footnotesize
Same as Fig.~\ref{fig2}
for a core less dense and less centrally concentrated than L1544, which 
may represent an earlier stage in the evolution towards the formation 
of a star.
The density drops as $r^{-0.4}$ inside 4000 AU, and as $r^{-2}$ between 
4000 and 15000 AU (the assumed cloud radius).  The central density is  
10$^5$ \percc \ (see right panel).  In this case, no ``holes'' are present
in the molecular distribution.  Although CO is not significantly depleted, 
\HTHOP  is still the main molecular ion at $r$ $\leq$ 6000 AU. 
Deuterium fractionation in \HCOP \ and \NTHP \
is constant across the cloud because of the negligible depletion of 
neutral species.
}
\label{fig4}
\end{figure}

In this case, we use the ``typical'' radial profile of a pre--stellar core 
deduced from millimeter observations of dust emission (Ward--Thompson et al. 
1994; Andr\'e et al. 2000; see Sect.~\ref{s-grad}), with a density profile 
$n(r)$ $\propto$ $r^{-0.4}$
inside 4000 AU and $n(r)$ $\propto$ $r^{-2}$ at larger radii.  The central
density, assumed constant inside 1000 AU, is 10$^5$ \percc (see right
panel in Fig.~\ref{fig4}).  This 
profile is similar to four starless cores recently 
studied by Tafalla et al. (2001), and thus it is representative 
of starless low mass cores (e.g. Benson \& Myers 1989).  
Indeed, L1544 is an exceptional starless low mass core, because 
of its large central densities and steep density profiles compared 
to the majority of the low mass core sample (see also Lee et al. 2001).
 The observed \CSEO \ column density towards the central
position is assumed 1$\times$10$^{14}$ \cmsq , or six times
lower than towards the L1544 molecular peak, based on Ladd et al.
(1998) and Tafalla et al. (2001) estimates in similar objects.

The input parameters for the ``flattened'' core are the same as in the 
L1544 model and  radial abundance profiles are displayed in 
Fig.~\ref{fig4}.
The first thing to note is that depletion factors are significantly smaller
than in L1544, reaching a maximum of 100 at 1000 AU.  \HCOP \ and \DCOP \ 
abundances show a shallower drop, and deuterium fractionation is 
not as effective as in L1544, because of the larger amount of 
CO in the gas phase (Sect.~\ref{s-depl}).  Once again, \HTHOP \ is the 
main molecular ion across the core.  

The electron 
fraction, inside the flattened region ($r \leq$ 4000 AU) is about 
10$^{-8}$, about a factor of 4 lower than predicted by the
``standard'' $x(e)-n(\MOLH )$ relation (see Sect.~\ref{s-model}).
This value of $x(e)$ implies a time--scale for the ambipolar diffusion
process of $t_{\rm ad}$ $\simeq$ 2.5$\times$10$^5$ yr, a factor of two 
larger than the free--fall time scale $t_{\rm ff}$ 
at the density of the flattened region (10$^5$ 
\percc ).  A consequence of this result is that, inside $r_f$, magnetic
fields and ion--neutral coupling are only marginally supporting the core
against gravity, so that if no other support is available, the core life
time is comparable to its dynamical time scale.  This is in contrast with 
the ``standard'' model of low mass star formation, where 
$t_{\rm ad}$/$t_{\rm ff}$ $\sim$ 10 (deduced from the ``standard'' ionization
model of molecular clouds) is used to justify the quasistatic treatment
of ambipolar diffusion in the formation of low--mass cores (Shu et al. 1987;
Mouschovias 1987). As discussed in Sect.~\ref{s-model}, depletion 
of metals is the cause of the reduced ionization degree.

\subsubsection{Column density profiles}

\begin{figure}{}
%\vspace*{-2cm}
\centerline{\includegraphics[angle=-90,width=15.0cm]{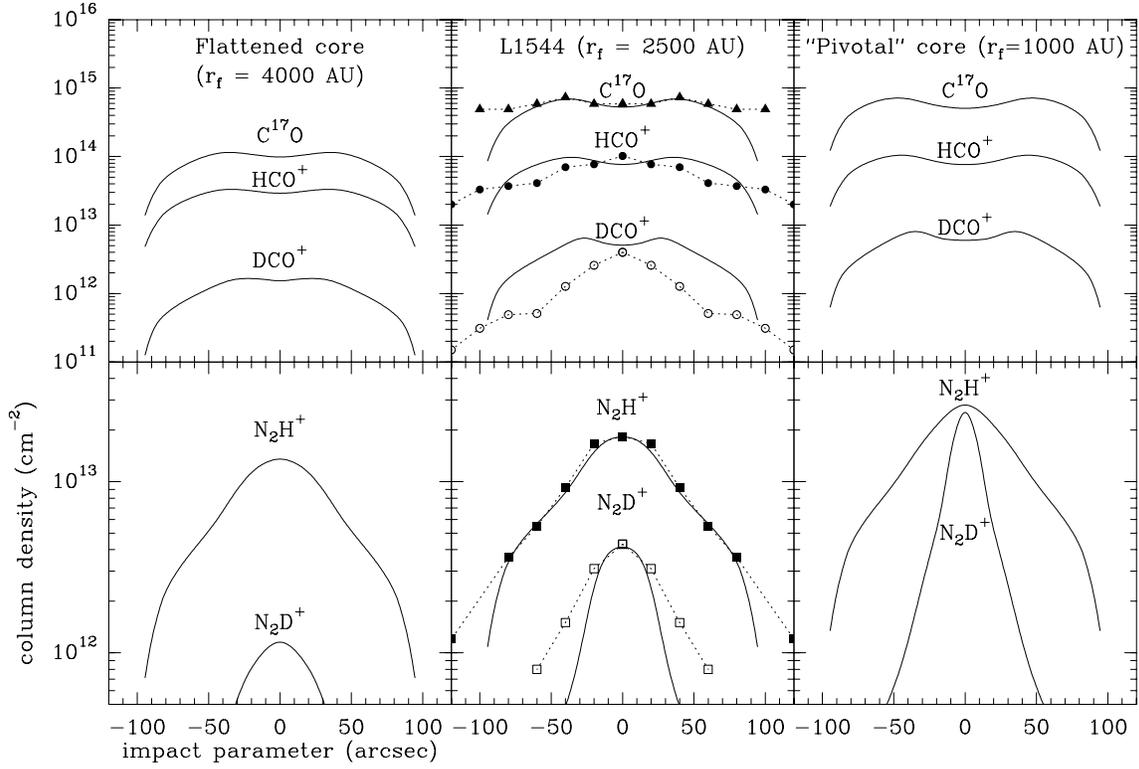}}
%\resizebox{10cm}{!}{\includegraphics{integrated_sminiato.ps}}
\vspace*{-1.5cm}
\caption{
\footnotesize
Column density profiles predicted in different dense clouds:
({\it left}) a flattened cloud (Sect.~\ref{s-flat}), representative of 
low mass cores; ({\it center}) 
L1544 (Sect.~\ref{s-l1544}) , a low mass starless core
on the verge of forming a star. Symbols are observed column densities 
averaged inside the corresponding bin (Caselli et al. 2001b); ({\it right})
a ``pivotal'' cloud (Sect.~\ref{s-spiky}), representative of more massive cores
($M$ $\sim$ 10 M$_{\odot}$), with higher central density and more 
centrally concentrated than L1544.  The $N(\NTDP )/N(\NTHP )$ column 
density ratio increases with the amount of CO depletion and, probably,
the core evolution (see text).  Column density profiles have been convolved 
with the 30m beam, assuming observations of the following transitions:
J = 1$\rightarrow$0 for \CSEO , \HCOP , \DCOP , and \NTHP ; J = 2$\rightarrow$1
for \DCOP and \NTDP .
}
\label{fig5}
\end{figure}

Integrating molecular abundances along the line of sight and 
convolving with the telescope beam, we can 
make predictions about column densities and molecular distributions 
of observable species in the above two model clouds.  
Fig.~\ref{fig5} shows the column density profiles of the ``flattened'' and 
``pivotal'' clouds, together with the L1544 results described in Caselli 
et al. 2001b (assuming the beam of the 30m antenna). 
The L1544 panels also report observational
data (symbols connected by dotted lines) which have been used to 
put constraints on the chemical model and find the best fitting set 
of parameters (see Caselli et al. 2001b and Sect.~\ref{s-model} for 
more details).  We note that the central ``valley'' in the \CSEO \ column
density profile 
becomes more pronounced in the pivotal core, 
because of the larger amount of
CO depletion.  The pivotal core also presents a well defined central depression
in the \HCOP \ and \DCOP \ column density profiles, accompanied by 
a small amount of limb brightening (observed in the \CSEO (1--0) map 
of L1544; Caselli et al. 1999).  The abundance ``hole'' in \HCOP \ and 
\DCOP \ (see Fig.~\ref{fig3}) should thus be observable (at the 30m 
telescope) in more evolved and/or
more massive cores.  

An interesting result of Fig.~\ref{fig5} is 
the difference in the \NTDP \ column density and \NTDP /\NTHP column 
density ratio between the three clouds, whereas $N(\DCOP )/N(\HCOP )$
does not significantly change.  The flattened core presents 
the smallest deuterium enhancement in \NTHP \  ($N(\NTDP )/N(\NTHP )$ = 
0.085, at the \NTHP \ column density peak).  Indeed, recent observations
toward four low mass starless cores
(Caselli et al. 2001, in prep.) show marginal detections of \NTDP \ and 
significantly lower amounts of deuterium fractionation compared to L1544.
On the other hand, the pivotal core shows a very sharp \NTDP \ morphology 
and a central $N(\NTDP )/N(\NTHP )$ $\simeq$ 1.  Thus, the 
$N(\NTDP )/N(\NTHP )$ column density ratio may be a good indicator of 
core evolution.

\section{Discussion and conclusions}
\label{s-disc}

Deuterium fractionation in abundant molecular ions offers a powerful tool 
to determine the fractional ionization of molecular clouds.  Simple 
chemical models of homogeneous clouds clearly show the strong
relation between the $N(\DCOP )/N(\HCOP )$ column density ratio and $x(e)$ 
(e.g. eqn. (\ref{exe})). Therefore, accurate estimates of $N(\DCOP )$ and 
$N(\HCOP )$ are the first steps towards an educated $x(e)$ guess.  This 
implies accounting for the optical depth of observed lines and self--absorption
effects, which can be accomplished via observations of rare isotopomers
(e.g. \DTHCOP , \HCEIOP , \HCSEOP ; Caselli et al. 2001b) and/or lines with 
hyperfine structure (e.g. the  \NTHP \ and \NTDP \ rotational lines; 
Caselli et al. 1995, Gerin et al. 2001).  However, the
degree of deuterium fractionation is also strongly affected by the 
depletion of neutral species (mainly CO and O; see Sect.~\ref{s-depl}), 
so that it is necessary to estimate the amount of CO depletion before
attempting $x(e)$ measurements.  Using the above observational data 
as constraints for detailed chemical models of homogeneous clouds, 
one can determine an average value of $x(e)$ (Sect.~\ref{s-detailed}). 
Typical ionization fractions are (i) between 10$^{-8}$ and 
10$^{-6}$ in low mass cores (Caselli et al. 1998), (ii)  
10$^{-7.3}$ $\la$ $x(e)$ $\la$ 10$^{-6.9}$ in more massive cores
(Bergin et al. 1999),  and (iii) $<$ 10$^{-8}$ in very massive massive 
clouds (de Boisanger et al. 1996).

The study of spatial variations in the ionization degree of a dense cloud
requires (i) maps of the dust continuum emission (or the dust extinction; e.g. 
Lada et al. 1994; Alves et al. 2001), to derive the density profile,  
and (ii) several molecular line maps
(CO, \HCOP , \DCOP , \NTHP , \NTDP ), to put constraints on the chemistry. 
In fact, as established from the recent study of L1544 
(Caselli et al. 2001a, 2001b) and other starless cores (Tafalla et al. 2001), 
dense cloud cores experience a phase of strong molecular abundance gradients 
which can be reproduced by chemical models only by taking into account
the density structure, gas phase depletion, and different binding energies 
of neutral species onto dust grain surfaces (see also Aikawa et al. 2001).  
In the case of L1544, CO 
is highly depleted at densities above about 10$^5$ \percc , causing 
the abundance of related species such as \HCOP \ and \DCOP \ to rapidly drop 
in the inner parts of the core. This explains the relatively low 
$N(\DCOP )/N(\HCOP )$ column density ratio ($\simeq$ 0.04) observed towards 
the L1544 dust emission peak.  On the other hand, the more volatile \MOLN \ 
maintains a rich nitrogen chemistry in the gas phase, causing a large deuterium
enhancement in \NTHP ($N(\NTDP )/N(\NTHP )$ $\simeq$ 0.24, toward the
same position).  Given that \MOLN \ is also precursor of \AMM , large ammonia 
deuteratium fractionations are expected, and indeed observed 
by Shah \& Wootten (2001) in L1544 and by Tin\'e et al. (2000) in similar
cores.  

Inside the ``flattened'' zone of L1544 ($r_f$ = 2500 AU), 
where the density is  $\sim$ 10$^6$ \percc , we derive an electron fraction of 
a few times 10$^{-9}$, implying an ambipolar diffusion time scale locally
similar to the free--fall rate, consistent with the idea that the L1544 core 
is close to dynamically collapse.  This is in agreement with the conclusions
of Aikawa et al. (2001) who best reproduce observed molecular distributions
using {\it fast} collapse in their dynamical--chemical model.  
The value we derive for $x(e)$ is 
about an order of magnitude smaller than the one deduced from the 
standard relation between $x(e)$ and $n(\MOLH )$, expected if the 
electron fraction is due to cosmic ray ionization alone and
without depletion (as in McKee 1989).  

Our chemical model applied to different types of clouds, predicts 
an observable increase in the $N(\NTDP )/N(\NTHP )$ column density ratio 
as the core evolves towards steeper density profiles, larger central 
densities,  and more CO--depleted inner regions ($N(\NTDP )/N(\NTHP )$ =
0.085, 0.23, and $\sim$ 1 for the flattened core, L1544, and the pivotal core,
respectively).  On the other hand, the $N(\DCOP )/N(\HCOP )$ ratio 
does not significantly change ( = 0.05, 0.07, and 0.08 for the flattened, 
L1544, and pivotal cores, respectively).  The different deuterium enhancements
in \HCOP \ and \NTHP \ reflect the cloud density structures, the molecular 
abundance gradients, and differential depletion of neutral species onto 
dust grains.  \HCOP \ and \DCOP \ do not trace the 
inner core regions, and, although the observed $N(\DCOP )/N(\HCOP )$ column 
density ratio is useful to put constraints on chemical models, the 
determination of the ionization degree inside the cloud requires  
observations of \NTHP \ and \NTDP .  These two species (in particular 
\NTDP ) maintain large abundances at densities $\ga$ 10$^6$ \percc \ so that
they offer a guide to chemical and physical properties, as well 
as the dynamical behaviour (see Caselli et al. 2001a), of the high density 
core nucleus, where the formation of a star will eventually take place.
This is extremely important for unveiling the initial conditions in the 
star formation process, including the fundamental parameter $x(e)$.   

%The presence of hyperfine structure 
%in the latter species (e.g. Caselli et al. 1995; Gerin et al. 2001),
%ensures accurate column density estimates 

\noindent
{\bf Acknowledgements}

I am grateful to all my collaborators and, in particular, to Malcolm Walmsley 
for stimulating discussions.  I thank the referees for their useful comments 
and suggestions, and I acknowledge support from ASI Grant 98-116 as well as 
from the MURST project ``Dust and Molecules in Astrophysical Environments''. 

\noindent
{\bf References}

\bib  Aikawa, Y., Ohashi, N., Inutsuka, S., Herbst, E., Takakuwa, S., 2001. 
Molecular evolution in collapsing prestellar cores. ApJ. 552, 639B653.

\bib Alves, J.F., Lada, C.J., Lada, E.A., 2001.  Internal structure of a cold dark 
molecular cloud inferred from the extinction of background starlight. Nature. 409,
159B161. 

\bib Anderson, I.M., Caselli P., Haikala, L.K., Harju, J., 1999.  Deuterium 
fractionation and the degree of ionization in the R Coronae Australis 
molecular cloud core. A\&A. 347, 983B999.

\bib Andr\'e P., Ward--Thompson, D., Motte, F., 1996. Probing the initial 
conditions of star formation: the structure of the prestellar core L 1689B. 
A\&A. 314, 625B635. 

\bib Benson, P.J., Myers, P.C., 1989. A survey for dense cores in dark clouds.
ApJS. 71, 89B108.

\bib Bergin, E.A., Langer, W.D., 1997. Chemical evolution in preprotostellar and 
protostellar cores. ApJ. 486, 316B328.

\bib Bergin, E.A., Plume, R., Williams, J.P., Myers, P.C., 1999. The Ionization 
Fraction in Dense Molecular Gas. II. Massive Cores. ApJ. 512, 724B739.

\bib Burke, J.R., Hollenbach, D.J., 1983. The gas-grain interaction in the 
interstellar medium - Thermal accommodation and trapping. ApJ. 265, 223B234.

\bib Butner, H.M., Lada, E.A., Loren, R.B., 1995. Physical properties of dense 
cores: \DCOP \ observations. ApJ. 448, 207B225.

\bib Caselli, P., 2000. The fractional ionization in molecular cloud cores.
In: Minh, Y.C., van Dishoeck, E.F. (Eds.), Astrochemistry: From Molecular 
Clouds to Planetary Systems, Proceedings of IAU Symposium 197. Astronomical 
Society of the Pacific, pp. 41B50.

\bib Caselli, P., Myers, P.C., Thaddeus, P., 1995. Radio-astronomical 
spectroscopy of the hyperfine structure of \NTHP . ApJ. 455, L77BL80.

\bib Caselli, P., Walmsley, C.M., 2000. Ionization and chemistry in the dense 
interstellar medium. In: Montmerle, T., Andr\'e, P. (Eds.), From Darkness
to Light, ASP Conference Series, in press.

\bib Caselli, P., Walmsley, C.M., Tafalla, M., Dore, L., Myers, P.C., 1999.
CO depletion in the starless cloud core L1544. ApJ. 523, L165BL169.

\bib Caselli, P., Walmsley, C.M., Zucconi, A., Tafalla, M., Dore, L., Myers, 
P.C., 2001a. Molecular ions in L1544. I. Kinematics. ApJ. in press. 

\bib Caselli, P., Walmsley, C.M., Zucconi, A., Tafalla, M., Dore, L., Myers, 
P.C., 2001b. Molecular ions in L1544. II. The ionization degree. ApJ. in press. 

\bib Caselli, P., Walmsley, C.M., Terzieva, R., Herbst, E., 1998. The 
ionization fraction in dense cloud cores. ApJ. 499, 234B249.

\bib Caux, E., Ceccarelli, C., Castets, A., Vastel, C., Liseau, R., 
Molinari, S., Nisini, B., Saraceno, P., White, G. J., 1999. Large atomic 
oxygen abundance towards the molecular cloud L1689N. A\&A. 347, L1-L4.

\bib Chiar, J.E., Adamson, A.J., Kerr, T.H., Whittet, D.C.B., 1995. High-resolution 
studies of solid CO in the Taurus dark cloud: characterizing the ices in quiescent
clouds. ApJ. 455, 234B243.

\bib Dalgarno, A., Lepp, S., 1984. Deuterium fractionation mechanisms in 
interstellar clouds. ApJ. 287, L47BL50. 

\bib de Boisanger, C., Helmich, F.P., van Dishoeck, E.F., 1996. The ionization 
fraction in dense clouds. A\&A. 310, 315B327.

\bib Draine, B.T., Sutin, B., 1987. Collisional charging of interstellar 
grains. ApJ. 320, 803B817. 

\bib Evans, N.J., II, Rawlings, J.M.C., Shirley, Y.L., Mundy, L.G., 2001. Tracing 
the mass during low-mass star formation. II. Modeling the submillimeter emission 
from preprotostellar cores. ApJ. 557, 193B208.

\bib Frerking, M.A., Langer, W.D., Wilson, R.W., 1982. The relationship 
between carbon monoxide abundance and visual extinction in interstellar clouds.
ApJ. 262, 590B605. 

\bib Fukui, Y., Hayakawa, S., 1981. Interactions of cosmic rays with molecular 
clouds. In: International Cosmic Ray Conference, 17th, Conference Papers. Volume 2.
Gif-sur-Yvette, Essonne, France, Commissariat a l'Energie Atomique, pp. 226B228.

\bib Gerin, M., Pearson, J.C., Roueff, E., Falgarone, E., Phillips, T.G., 
2001. Determination of the hyperfine structure of \NTDP . ApJ. 551, L193BL197.

\bib Goicoechea, J.R., Cernicharo, J., 2001. Far-Infrared detection of \HTHOP 
\ in Sagittarius B2. ApJ. 554, L213BL216.

\bib Graedel, T.E., Langer, W.D., Frerking, M.A., 1982. The kinetic chemistry of 
dense interstellar clouds. ApJS. 48, 321B368.

\bib Guel\'in, M., Langer, W.D., Wilson, R.W., 1982. The state of ionization in 
dense molecular clouds. A\&A. 107, 107B127. 

\bib Hasegawa, T.I., Herbst, E., 1993. New gas--grain chemical models of quiescent 
dense interstellar clouds - The effects of \MOLH \ tunnelling reactions and cosmic 
ray induced desorption. MNRAS. 261, 83B102.

\bib Hartquist, T.W., Morfill, G.E., 1983. Evidence for the stochastic acceleration
of cosmic rays in supernova remnants. ApJ. 266, 271B275.

\bib  Hartquist, T.W., Morfill, G.E., 1994. Cosmic ray diffusion at energies of 
1 MeV to 10$^5$ GeV. Ap\&SS. 216, 223B234.

\bib Herbst, E., Leung, C.M., 1989. Gas--phase production of complex hydrocarbons, 
cyanopolyynes, and related compounds in dense interstellar clouds. ApJS. 69, 
271B300.

\bib Herbst, E., Klemperer, W., 1973. The formation and depletion of 
molecules in dense interstellar clouds. ApJ. 185, 505B534.

\bib Jones, A.P., Williams, D.A., 1985. Time-dependent sticking coefficients and 
mantle growth on interstellar grains. MNRAS. 217, 413B421.

\bib Kokoouline, V., Greene, C.H., Esry, B.D., 2001.  Mechanism for the 
destruction of \HTHP \ ions by electon impact. Nature. 412, 891B894.

\bib Lada, C.J., Lada, E.A., Clemens, D.P., Bally, J., 1994. Dust extinction and 
molecular gas in the dark cloud IC 5146. ApJ. 429, 694B709. 

\bib Ladd, E.F., Fuller, G.A., Deane, J.R., 1998. \CEIO \ and \CSEO \ observations 
of embedded young stars in the Taurus Molecular Cloud. I. Integrated intensities 
and column densities. ApJ. 495, 871B890.

\bib Larsson, M., Lepp, S., Dalgarno, A., et al., 1996. Dissociative recombination 
of \HTDP  and the cosmic abundance of deuterium. A\&A. 309, L1BL3.

\bib Lee, H.-H., Bettens, R.P.A., Herbst, E., 1996. Fractional abundances of 
molecules in dense interstellar clouds: A compendium of recent model results.
A\&AS. 119, 111B114.

\bib Lee, C.W., Myers, P.C., Tafalla, M., 2001. A survey for infall motions toward 
starless cores. II. CS(2--1) and \NTHP (1--0) mapping observations. ApJS. in press.

\bib Linsky, J.L., Diplas, A., Wood, B.E., Brown, A., Ayres, T.R., Savage, 
B.D., 1995. ApJ. 451, 335B351.

\bib Mathis, J.S., Rumpl, W., Nordsieck, K.H., 1977. The size distribution of 
interstellar grains. ApJ. 217, 425B433.

\bib McCall, B.J., Geballe, T.R., Hinkle, K.H., Oka, T., 1999. Observations 
of H$^+_3$ in dense molecular clouds. ApJ. 522, 338B348.

\bib McKee, C.F., 1989. Photoionization--regulated star formation and the 
structure of molecular clouds. ApJ. 345, 782B801.

\bib McKee, C.F., Zweibel, E.G., Goodman, A.A., Heiles, C., 1993. Magnetic 
fields in star-forming regions - Theory. Protostars and Planets III Editors, 
Eugene H. Levy, Jonathan I. Lunine; Publisher, University of Arizona Press, 
Tucson, Arizona. 327B366.

\bib Mestel, L., Spitzer, L., Jr., 1956. Star formation in magnetic dust 
clouds. MNRAS. 116, 503B514.

\bib Millar, T.J., Bennett, A., Herbst, E., 1989. Deuterium fractionation in 
dense interstellar clouds. ApJ. 340, 906B920. 

\bib Morton, D.C., 1974. Interstellar abundances toward zeta Ophiuchi. ApJ.
193, L35-L39.

\bib Motte, F., Andr\'e, P., 2001. The circumstellar environment of low-mass 
protostars: A millimeter continuum mapping survey. A\&A. 365, 440B464.

\bib Mouschovias, T.Ch., 1987. Star formation in magnetic interstellar clouds. 
I - Interplay between theory and observations. II - Basic theory. In: Morfill, 
G.E.,  Scholer, M. (Eds.), Physical Processes in Interstellar Clouds. Dordrecht,
Reidel. pp. 453B489.

\bib Ohashi, N., Lee, S.W., Wilner, D.J., Hayashi, M., 1999. CCS imaging of the 
starless core L1544: an envelope with infall and rotation. ApJ. 518, L41BL44.

\bib Oppenheimer, M., Dalgarno, A., 1974. The fractional ionization in dense 
interstellar clouds. ApJ. 192, 29B32.

\bib Phillips, T.G., van Dishoeck, E.F., Keene, J., 1992. Interstellar 
\HTHOP \ and its relation to the O$_2$ and H$_2$O abundances. ApJ. 399, 
533B550.

\bib Rodgers, S.D., Charnley, S.B., 2001. Gas--phase production of 
NHD$_2$ in L134N. ApJ. 553, 613B617. 

\bib Sadlej, J., Rowland, B., Devlin, J.P., Buch, V., 1995.
J. Chem. Phys. 102, 4804

\bib Shah, R.Y., Wootten, A., 2001. Deuterated ammonia in galactic protostellar 
cores. ApJ. 554, 933B947.

\bib Shu, F.H., 1977.  Self--similar collapse of isothermal spheres and star 
formation. ApJ. 214, 488--497.

\bib Shu, F.H., Adams, F.C., Lizano, S., 1987. Star formation in molecular 
clouds - Observation and theory. ARA\&A. 25, 23B81.

\bib Spitzer, L., Jr., 1978, Physical Processes in the Interstellar Medium. 
New York, Wiley.

\bib Tafalla, M., Mardones, D., Myers, P.C., Caselli, P., Bachiller, R., Benson, 
P.J., 1998. L1544: A starless dense core with extended inward motions. ApJ.
504, 900B914.

\bib Tafalla, M., Myers, P.C., Caselli, P., Walmsley, C.M., Comito, C., 2001.
Sistematic molecular differentiation in starless cores. ApJ. submitted.

\bib  Tielens, A.G.G.M., Allamandola, L.J., 1987. Evolution of interstellar dust.
In: Hollenbach, D.J., Thronson, H.A., Jr. (Eds.), Interstellar Processes. Dordrecht,
Kluwer, pp. 333B376.

\bib Tin\', S., Roueff, E., Falgarone, E., Gerin, M., Pineau des For\^ets, G., 
2000. Deuterium fractionation in dense ammonia cores. A\&A. 356, 1039B1049.

\bib Umebayashi, T., Nakano, T., 1990. Magnetic flux loss from interstellar clouds.
MNRAS. 243, 103B113.

\bib van Dishoeck, E.F., Blake, G.A., Draine, B.T., Lunine, J.I., 1993. 
The chemical evolution of protostellar and protoplanetary matter. In: 
Levi, E.H., Lunine, J.I. (Eds.), Protostars and planets III. Tucson, 
Univ. Arizona Press, pp. 163B241.

\bib Ward-Thompson, D., Motte, F., Andr\'e, P., 1999. The initial conditions of 
isolated star formation - III. Millimetre continuum mapping of pre-stellar cores.
MNRAS. 305, 143B150. 

\bib Ward--Thompson, D., Scott, P.F., Hills, R.E., Andr\'e, P., 1994. A 
submillimetre continuum survey of pre protostellar cores. MNRAS. 268, 276B290.

\bib Watson, W.D., 1976. Interstellar molecule reactions. Rev. Mod. Phys. 48,
513B550.

\bib Willacy, K., Langer, W.D., Velusamy, T., 1998. Dust emission and molecular 
depletion in L1498. ApJ. 507, L171BL175.

\bib Williams, J.P., Bergin, E.A., Caselli, P., Myers, P.C., Plume, R., 1998.
The ionization fraction in dense molecular gas. I. Low-mass cores. ApJ. 503,
689B699.

\bib Williams, J.P., Myers, P.C., Wilner, D.J., di Francesco, J., 1999. A 
high--rsolution study of the slowly contracting, starless core L1544. ApJ. 513,
L61BL64.

\bib Wootten, A., Snell, R., Glassgold, A.E., 1979. The determination of 
electron abundances in interstellar clouds. ApJ. 234, 876B880.

\bib Zucconi, A., Walmsley, C.M., Galli, D., 2001. The dust temperature 
distribution in prestellar cores. A\&A. 376, 650B662.

\newpage

%Acknowledgements
%References
%Figure captions
%Figures

\end{document}